\documentclass[
aps,
twocolumn,
prb,
longbibliography,
groupedaddress,
superscriptaddress,
showpacs,
]
{revtex4-1}
\RequirePackage{amsmath,amssymb,bm,wasysym}
\RequirePackage{graphicx}
\RequirePackage{hyperref}
\usepackage{siunitx}
\usepackage{tikz}
\hypersetup{
breaklinks={true},
citecolor={blue},
colorlinks={true},
linkcolor={blue},
pdfcreator={\LaTeX and\flqq hyperref\frqq},
pdffitwindow={true},
pdfmenubar={true},
pdfpagelayout={OneColumn},
pdfstartview={FitH},
pdftoolbar={true},
plainpages={false},
pdfpagemode={UseThumbs},
}
\usepackage{braket}
\DeclareMathAlphabet{\mathpzc}{OT1}{pzc}{m}{it}
\renewcommand{\vec}[1]{\boldsymbol{#1}}

\newcommand{\bea}{\begin{eqnarray*}}
\newcommand{\eea}{\end{eqnarray*}}
\newcommand{\bne}{\begin{equation*}}
\newcommand{\ede}{\end{equation*}}

\newcommand{\bnen}{\begin{equation}}
\newcommand{\eden}{\end{equation}}
\newcommand{\bean}{\begin{eqnarray}}
\newcommand{\eean}{\end{eqnarray}}
\newcommand{\bsen}{\begin{subequations}}
\newcommand{\esen}{\end{subequations}}

\newcommand{\bna}{\begin{array}}
\newcommand{\eda}{\end{array}}
\newcommand{\bnm}{\begin{enumerate}}
\newcommand{\edm}{\end{enumerate}}
\newcommand{\bni}{\begin{itemize}}
\newcommand{\edi}{\end{itemize}}

\renewcommand{\vec}[1]{\text{\boldmath{$ #1 $}}}

\newcommand{\minus}{\scalebox{0.45}[1.0]{$-$}}

\begin{document}
\pagestyle{plain}
\title{Fast electron spin flips via strong subcycle electric excitation}
\author{M\'at\'e Tibor Veszeli}
\affiliation{Institute of Physics, E\"{o}tv\"{o}s University, 1518 Budapest, Hungary}
\author{Andr\'{a}s P\'{a}lyi}
\affiliation{
MTA-BME Exotic Quantum Phases "Momentum" Research Group and Department of Physics,
Budapest University of Technology and Economics, 1111 Budapest, Hungary}
\pacs{
}
\begin{abstract}
An important goal in quantum information processing
is to reduce the duration of quantum-logical operations.
Motivated by this, we provide a theoretical analysis
of electrically induced 
fast  dynamics of a single-electron spin-orbit qubit.
We study the example of a one-dimensional quantum dot
with Rashba spin-orbit interaction and
harmonic driving, and focus on the 
case of strong driving,  when the real-space 
oscillation amplitude of the driven electron is comparable to the width
of its wave function.
We provide simple approximate analytical 
relations between the qubit Larmor frequency, the
shortest achievable qubit-flip time, and the driving amplitude
required for the shortest achievable qubit flip.
We find that these relations compare well with results
obtained from numerical simulations of the qubit dynamics.
Based on our results, we discuss practical guidelines to
maximize speed and quality of electric single-qubit operations
on spin-orbit qubits.
\end{abstract}
\maketitle

\section{Introduction}
\label{sec:int}

An important physical quantity in quantum information processing
is the speed of single-qubit logical operations. 
Often the quality of a certain physical qubit realization is 
characterized by the ratio between the decoherence time
and the time required to perform coherent operations; 
the larger this ratio (quality factor), the better the qubit. 
Motivated by this, here we theoretically study a way to minimize
single-qubit gate times for single-electron spin-orbit qubits
in quantum dots 
 \cite{nadj-perge_spin-orbit_qubit, nowack_esr,nadj-perge,schroer_field_tuning,rui-you-sun-nori,romhanyi, golovach_edsr, flindt-sorensen}. 
More precisely, we investigate how spin-orbit-mediated
qubit-flip processes can 
be made fast using short pulses of intense ac electric fields,
with a pulse length comparable to the driving period.

\begin{figure}
 \centering
 \includegraphics[width=\columnwidth]{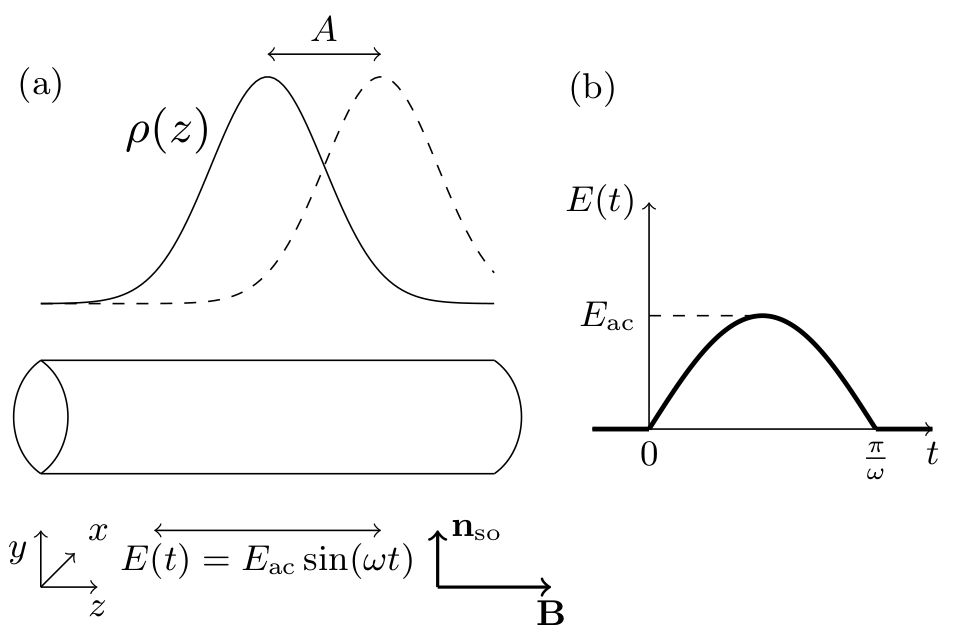}
 \caption{{\bf Half-cycle excitation of a one-dimensional
 spin-orbit qubit.
 } 
(a) A spin-orbit qubit is defined in a one-dimensional 
nanowire (cylinder), in the presence of a magnetic field
$\vec B$ and spin-orbit interaction, the latter characterized
by the spin-orbit field direction $\vec{n}_\text{so}$. 
In equilibrium, the electron occupies the ground state, 
with charge density $\rho(z)$. 
Upon electric excitation $E(t)$,
the electron is moved back and forth with amplitude
$A$ along the wire, 
inducing dynamics between the spin-orbit qubit basis
states.
(b) A half-cycle excitation: an appropriately tuned strong but short
half-cycle excitation pulse induces a fast flip of the spin-orbit
qubit.
 }
 \label{fig:setup}
\end{figure}

We will consider harmonic electric driving, described
by the Hamiltonian 
\begin{equation}
 \label{eq:electric}
H_E(t) = e E_\text{ac} z \sin(\omega t),
\end{equation}
where $e$ is the elementary charge, 
$E_\text{ac}$ is the amplitude of the driving electric field, 
$z$ is a spatial coordinate, and $\omega$ is the driving (angular)
frequency. 
We will consider the situation when the qubit is in 
its ground state at $t=0$, driving starts at $t=0$, and is stopped at
$t=\tau$; we call $\tau$ the pulse duration. 
A constraint we introduce is that
$\omega \tau = N\pi$; 
that is, the pulse duration is an integer number of 
half-cycles.
This constraint comes from a practical consideration: 
if driving is stopped at some arbitrary $\tau$, 
then the resulting electronic state at $t=\tau$ could be 
very different from the eigenstates of the static Hamiltonian, 
and therefore be subject to significant time evolution even after
the driving is stopped. 
We introduce the above constraint $\tau = N\pi/\omega$ 
to avoid such a complication. 
This also implies that the shortest pulse in this framework 
corresponds to a single half-cycle of the excitation, that is,
$N=1$ and pulse duration $\tau = \pi/\omega$, 
as shown in Fig~\ref{fig:setup}b.
This is the case we focus on. 

The main question we address in this work is the following. 
Assume that certain practical constraints regarding
the system parameters are given. 
Under those constraints, how fast can we perform a single
qubit flip? 
What is the corresponding driving strength $E_\text{ac}$ 
we have to apply to the system to achieve that 
fast qubit-flip? 
How do those quantities depend on the experimentally 
tunable parameters, such as the magnetic field strength and
the spin-orbit interaction strength?
We answer these questions within a specific model, 
and use those answers to establish practical guidelines 
for maximizing
 speed and quality of electric single-qubit operations
on spin-orbit qubits.

In particular, the key findings within our model are
as follows. 
(i) We estimate that a fast qubit flip can be performed 
within 10 ps using a strong half-cycle pulse, assuming that
the strong pulse does not ionize the quantum dot
(see Section \ref{sec:estimates}).
(ii) 
The qubit flip can be made faster by increasing the 
external magnetic field $B$, and 
correspondingly increasing the drive frequency $\omega$, 
without the need of increasing the drive strength $E_\text{ac}$
(see Section \ref{sec:estimates},
Section \ref{sec:results}, and Fig.~\ref{fig:max}).
(iii) Counterintuitively, decreasing the strength of the spin-orbit
interaction might allow for a faster qubit flip, although that
requires an increased drive strength
(see Section \ref{sec:discussion}).

\section{Setup}
\label{sec:setup}

We examine a one-dimensional quantum dot,
see Fig.~\ref{fig:setup}a, described by the Hamiltonian
\bean
\label{eq:hamiltonian}
H(t) = H_\text{osc} + 
H_\text{so} +
H_\text{Z} +
H_E(t).
\eean
Here, $H_\mathrm{osc}$ is the harmonic oscillator term, that, is we approximate the confining potential with a parabola:
 \begin{equation}
  H_\mathrm{osc} = \dfrac{ p_z^2 }{ 2m } + \dfrac{ m \omega_0 }{ 2 } z^2 = \hbar \omega_0 \left( a^\dag a + \dfrac{ 1 }{ 2 } \right),
 \end{equation}
where $p_z = -i \hbar \partial_z$, and $a, a^\dag$ are the ladder operators.
The width of the oscillator ground-state wave function 
is $L = \sqrt{\hbar / (m\omega_0)}$. 
The electron is in a z-directional 
homogeneous 
magnetic field, which causes a Zeeman splitting between the spin-up and spin-down states,
described by
 \begin{equation}
  H_\mathrm{Z} = -\dfrac{ g^* }{ 2 } \mu_\mathrm{B} B \sigma_z = - \dfrac{ 1 }{ 2 } \tilde{B} \sigma_z.
 \end{equation}
Here, $B$ is the magnetic field strength, $\mu_\mathrm{B}$ is the Bohr-magneton, $g^*$ is the effective g-factor and $\sigma_z$ is 
the third Pauli matrix. The magnetic field points to the $z$ direction. In this setup, electric spin control is enabled by spin-orbit coupling:
 \begin{equation}
  H_\mathrm{so} = 
  \alpha p_z \vec{n}_\text{so} \cdot \vec \sigma = 
  \alpha p_z \sigma_y = i \tilde{\alpha} \left( a^\dag - a \right) \sigma_y,
 \end{equation}
where $\alpha$ is the Rashba spin-orbit coupling strength. 
The direction of the spin-orbit field is given by the unit vector
$\vec{n}_\text{so}$, which we choose to be along 
the $y$ direction, as shown in Fig.~\ref{fig:setup}a.
The electric driving has already been introduced in Eq.~\eqref{eq:electric}. 
With ladder operators it is expressed as:
 \begin{equation}
  H_E(t)  = \tilde{E}_\mathrm{ac} \sin(\omega t)\left( a^\dag + a \right).
 \end{equation}
 Note that we will consider cases when this electric driving is
 adiabatic with respect to the oscillator level spacing, 
 $\omega \ll \omega_0$.
 In that case, the driving induces spatial oscillations of the 
 electronic wave packet with an approximate amplitude 
 $A = \frac{\sqrt{2} \tilde{E}_\text{ac}}{\hbar \omega_0} L$, 
 as shown in Fig.~\ref{fig:setup}a. 
 
The quantities with tilde all have energy dimension,
hence we can use them to compare the strengths of the
different terms of the Hamiltonian; their definitions
are
\begin{subequations}
 \begin{equation}
  \tilde{B} = g^{*} \mu_\mathrm{B} B,
 \end{equation}
 \begin{equation}
  \tilde{\alpha} = \alpha \sqrt{ \dfrac{ m \hbar \omega_0 }{ 2 } },
 \end{equation}
 \begin{equation}
  \tilde{E}_\mathrm{ac} = e E_\mathrm{ac} \sqrt{ \dfrac{ \hbar }{ 2 m \omega_0 } }.
 \end{equation}
\end{subequations}
The basis states of the spin-orbit qubit are the ground and first excited states of the static Hamiltonian $H_\mathrm{osc} + H_\mathrm{Z} + H_\mathrm{so}$, to be denoted by $\ket{\mathrm{g}}$ and $\ket{\mathrm{e}}$, respectively.

\section{Qubit dynamics at weak driving}
\label{sec:perturbative}

Here, we outline the derivation leading to the weak driving
results for the Larmor and the Rabi frequency
\cite{golovach_edsr, rui-you-sun-nori, romhanyi, flindt-sorensen}.
Even though these results are already known, we include them
here for self-containedness and to set the stage for our
further results.
The parameter range we consider here 
is $\tilde{\alpha} \lesssim \hbar \omega_0$,
$\tilde{B} \ll \hbar \omega_0$, $\tilde{E}_\mathrm{ac} \ll \hbar \omega_0$, 
and we refer to the last condition as the
\emph{weak-driving limit}.
In the first step, we perform a well-known unitary 
transformation\cite{LevitovRashba,rui-you-sun-nori}
that eliminates the spin-orbit term $H_\text{so}$,
but renders the homogeneous magnetic field
inhomogeneous. 
Second, we express the approximate energy eigenstates
of the resulting static Hamiltonian using first-order
perturbation theory in the magnetic field.
This allows us to identify the basis states of the 
spin-orbit qubit.
Then, we project the complete Hamiltonian, 
including the driving term $H_E(t)$,
to this qubit subspace, to reveal 
the qubit Larmor frequency where resonant excitation is expected, 
and the Rabi frequency of the qubit Rabi oscillations induced 
by the electric drive. 

The first step is to perform the unitary 
transformation\cite{LevitovRashba,rui-you-sun-nori}
\bean
\label{eq:levitovrashba}
U = e^{i m \alpha z \sigma_y/\hbar}
\eean
on the Hamiltonian of Eq.~\eqref{eq:hamiltonian}. 
This results in the transformed Hamiltonian
\bean
H'(t) = U H(t) U^\dag
= H'_\text{osc} + H'_\text{so} + H'_\text{Z} + H'_\text{E}(t),
\eean
containing the following terms:
\begin{subequations}
\label{eq:hlevitovrashba}
\bean
\label{eq:hlevitovrashba1}
H'_\text{osc} + H'_\text{so} &=& 
 U H_\text{osc} U^\dag + U H_\text{so} U^\dag
= H_\text{osc},
\\
H'_\text{Z} &=& 
U H_\text{Z} U^\dag \\ \nonumber
&=& - \frac 1 2 \tilde{B} \left[
\cos\left(\frac{2m\alpha z}{ \hbar}\right) \sigma_z
+\sin\left(\frac{2m\alpha z}{ \hbar}\right) \sigma_x
\right], \\
H'_E &=& U H_E U^\dag = H_E.
\eean
\end{subequations}
Hence the spin-orbit interaction is eliminated and
the magnetic field acquires a spiral-type spatial dependence.
Note that a position- and spin-independent constant has been 
omitted in Eq.~\eqref{eq:hlevitovrashba1}.

The second step is to express the two
lowest-energy
eigenstates 
$\ket{\mathrm{g'}}$ and $\ket{\mathrm{e'}}$
of the transformed static Hamiltonian,
using first-order perturbation theory in the Zeeman Hamiltonian.
Finally, we project the complete transformed
Hamiltonian to the qubit subspace
spanned by $\ket{\mathrm{g'}}$ and $\ket{\mathrm{e'}}$  using the projector 
$P= \ket{\mathrm{g'}}\bra{\mathrm{g'}} + 
\ket{\mathrm{e'}}\bra{\mathrm{e'}}$. 
This procedure yields
\bean
\label{eq:hqubit}
H'_\text{q}(t) = P H'(t) P = 
-\frac 1 2 \hbar \omega_L \tau_z 
- \hbar \Omega_\text{R} \tau_x \sin (\omega t),
\eean
where $\tau_i$ are the Pauli operators in the qubit subspace, 
e.g., $\tau_z = \ket{\mathrm{g'}}\bra{\mathrm{g'}} -
\ket{\mathrm{e'}}\bra{\mathrm{e'}}$.
After the rotating-wave approximation\cite{Shirley},
the Schr\"odinger equation corresponding to $H'_\text{q}$
can be solved exactly.
The Larmor frequency is
\newcommand{\teac}{\tilde E_\text{ac}}
\newcommand{\talpha}{\tilde \alpha}
\newcommand{\tb}{\tilde B}
\bean
\label{eq:larmor}
\hbar \omega_\text{L} = \tilde B 
e^{-2\left(\frac{\talpha}{\hbar \omega_0}\right)^2},
\eean
which describes the spin-orbit-induced suppression of the 
Zeeman splitting \cite{rui-you-sun-nori}, and the Rabi frequency
upon resonant driving $\omega = \omega_\mathrm{L}$ is
 \cite{golovach_edsr,rui-you-sun-nori,romhanyi}.
\bean
\label{eq:rabipert}
 \Omega_\text{R} = 
\frac{2 \teac \talpha  }{(\hbar \omega_0)^2} \omega_\text{L}.
\eean
In this resonant case the qubit can be flipped from its ground 
state to its excited state, so the excited-state population
$P_\mathrm{e}$ shows simple Rabi oscillations:
\bean
\label{eq:rabiosc}
P_\mathrm{e}(\tau) = \sin^2 \left(\Omega_\text{R} \tau / 2\right).
\eean

\section{Practical considerations and estimates}
\label{sec:estimates}

As stated in the introduction, the main questions we address
here are as follows. 
Given certain practical constraints, 
how fast can we perform a single qubit flip? 
What is the corresponding drive strength required for that
fast flip? 
How do these quantities depend on the tunable 
parameters? 

We can develop naive  answers to these questions
based on the analytical results \eqref{eq:larmor} and \eqref{eq:rabipert} 
obtained 
for the weak-driving limit.
As it was discussed in section \ref{sec:int}, the pulses have a 
duration $\tau=N \pi / \omega$, and in the resonant case is
\bean
\label{eq:tau}
\tau = N \pi/\omega_\text{L}.
\eean
The shortest pulse duration corresponds to a single
half-cycle, that is, to the $N=1$ case.
For a qubit flip, the driving amplitude should fulfill
$P_\text{e}(\tau) = 1$, that is, 
$ \Omega_\text{R} \tau/2  = \pi/2$.
From this condition, using Eq.~\eqref{eq:rabipert}, 
we obtain that the drive strength corresponding to the
shortest qubit-flip pulse is
\bean
\label{eq:naiveteac}
\teac = \frac{(\hbar \omega_0)^2}{ 2 n \tilde \alpha}.
\eean

It is important to note that the result \eqref{eq:naiveteac} is naive 
in the sense that the results \eqref{eq:rabipert} and
\eqref{eq:rabiosc} are established only in the weak-driving limit, $\teac \ll \hbar \omega_0$, 
whereas a parameter set fulfilling Eq.~\eqref{eq:naiveteac},
may lie outside this parameter regime.
For example, 
if $N=1$, $\tilde \alpha / \hbar \omega_0 = 0.5$, then
the drive strength from Eq.~\eqref{eq:naiveteac}
is $\teac = \hbar \omega_0$
which contradicts the weak-driving condition
$\teac \ll \hbar \omega_0$. 
(Note that this choice of $\tilde{\alpha} / \hbar \omega_0$ 
maximizes the Rabi frequency as the function
of spin-orbit strength, see 
Eqs. \eqref{eq:larmor} and \eqref{eq:rabipert}.)

In the rest of this work, we study the qubit-flip processes
induced by strong driving, in the range
$\teac \lesssim \hbar \omega_0$.
Surprisingly, as we show in Section \ref{sec:strong_driving}, 
we find that 
the characteristics of the fast qubit-flips induced by half-cycle
driving are rather well described by  naively extrapolating 
the results \eqref{eq:tau} and \eqref{eq:naiveteac}
obtained for the weakly driven case.
First we present an analytical derivation, then we compare it
to the numerical simulations in Sec.~\ref{sec:strong_driving}
and in Sec.~\ref{sec:results}.

Anticipating that the extrapolation of the weak-driving results
gives an accurate description of the strongly driven case,
here we use it to draw practical conclusions.
If our goal is to achieve qubit-flips as fast as possible, i.e., 
to minimize $\tau$, then Eq.~\eqref{eq:tau}
advises to set the Larmor frequency as high as possible. 
Practically, one upper bound on the Larmor frequency is
given by the maximal achievable magnetic field strength. 
Another upper bound follows from the resonant-driving
condition: the maximal Larmor frequency is limited
by the maximum available driving frequency.

For the sake of a practical example, 
assume that the maximum available driving frequency 
is $\omega_\text{max}/(2\pi) = 50\, \text{GHz}$.
Furthermore, 
consider an InAs nanowire quantum
dot\cite{schroer_field_tuning,nadj-perge_spin-orbit_qubit}
with
electronic effective mass 
 $m=0.023 m_0$,
effective g-factor
$g^*=10$,
orbital level spacing
 $\hbar \omega_0 = 1 \, \text{meV}$,
and  spin-orbit strength
$\alpha= 62 \, \text{km}/\text{s}$.
These parameters imply an electronic confinement length
$L = \sqrt{\hbar/(m \omega_0)} \approx 58\, \text{nm}$.

In this case, 
the magnetic field strength corresponding to 
$\omega_\text{L} = \omega_\text{max}$ is
$B\approx 0.6\, \text{T}$, which is experimentally feasible.
Hence, in this case, the component limiting the achievable
qubit-flip speed is the limited driving frequency. 
In this example, Eq.~\eqref{eq:tau} predicts that 
the shortest achievable qubit-flip takes
$\tau = 10\, \text{ps}$, which is 3 orders of magnitude shorter 
than the $10$ ns 
qubit-flip time scale demonstrated experimentally in 
Ref.~\onlinecite{nadj-perge_spin-orbit_qubit}.
To achieve this, the driving amplitude should be set to
$\teac = \hbar \omega_0$, 
corresponding to a driving field amplitude 
$E_\text{ac} \approx 25 \, \text{kV}/\text{m}$, and
an electron oscillation amplitude 
$A = \frac{\sqrt{2} \tilde{E}_\text{ac}}{\hbar \omega_0} L \approx 82 \, \text{nm} $.

\section{Strong driving described in the co-moving frame}
\label{sec:strong_driving}

Having established the weakly driven spin dynamics
in Sec.~\ref{sec:perturbative},
it is natural to ask how this dynamics changes if
the drive strength is not perturbative, i.e.,
if we do not assume 
$\tilde{E}_\text{ac} \ll \hbar \omega_0$ 
but allow for $\tilde{E}_\text{ac} \lesssim \hbar \omega_0$
instead?

As we already noted above, the analytical results 
Eqs.~\eqref{eq:larmor} and \eqref{eq:rabipert} 
and the corresponding dynamics is recovered to a large
extent in the strong-driving case.
Our strategy to obtain an analytical description of the 
strongly driven qubit dynamics is as follows.
As the first step, following the perturbative
case in Section \ref{sec:perturbative},
we apply the transformation $U$ of
Eq.~\eqref{eq:levitovrashba} on our Hamiltonian $H$
to eliminate spin-orbit interaction at the price of
rendering the magnetic field inhomogeneous.
Then, we transform to a spatial reference frame that is 
co-moving with the center of the time-dependent
harmonic oscillator potential $\frac{1}{2} m \omega_0^2 z^2 + 
e E_\text{ac} z \sin(\omega t)$,
and to a spin reference frame which is co-moving with
the local magnetic field at the center of the confinement potential.
These transformations then allow us to focus again on a 
two-dimensional low-energy effective qubit subspace that
is sufficiently decoupled from higher-lying energy 
eigenstates, and thereby to deduce analytical results for the
qubit dynamics. 

The first transformation, $U$ results in 
the Hamiltonian terms collected in 
Eq.~\eqref{eq:hlevitovrashba}.
The second transformation, 
    i.e., the spatial transformation to the co-moving frame
    \cite{PhysRevLett.99.246601, 0268-1242-24-6-064004, PhysRevB.89.115409},
    is described by the time-dependent unitary operator
    \bean
    \label{eq:transform2}
    W(t) = \sum_{n  = 0 }^{\infty} \ket{n} \bra{n(t)},
    \eean
    where $\ket{n}$ is the $n$th eigenstate of the 
    undriven harmonic oscillator $H_\text{osc}$, 
    and $\ket{n(t)}$ is the instantaneous eigenstate of the 
    driven harmonic oscillator $H_\text{osc} + H_E(t)$. 
    Note that an alternative way to express this
    transformation is 
    \bean
    W(t) = e^{-i A \sin (\omega t) p_z/\hbar}.
    \eean
    This time-dependent unitary transformation 
    results in the transformed Hamiltonian
    \bean
    H''(t) &=& W(t) H'(t) W^\dag(t) + i \hbar \dot{W}(t) W^\dag(t)
    \nonumber  \\ 
    &=& H_\text{osc} + H'_\text{Z}(z-A\sin \omega t)
    + A \omega \cos (\omega t) p_z. \label{eq:hcomoving}
    \eean

    The first two terms of Eq.~\eqref{eq:hcomoving}
    describes an electron in a static harmonic
    confinement subject to an inhomogeneous spiral-like, oscillating 
    Zeeman field.
    As the final, third transformation, we eliminate the 
    time dependence of the inhomogeneous Zeeman field via
    the time-dependent spin-rotating unitary transformation
    \bean
    \label{eq:transform3}
    S(t) = e^{-i \frac{m \alpha A \sin(\omega t)}{\hbar} \sigma_y}.
    \eean
This yields
\bean
H'''(t) &=& S(t) H''(t) S^\dag(t) + i \hbar \dot{S}(t) S^\dag(t) 
\nonumber
\\
&=& H_\text{osc} + H'_\text{Z}(z) + A \omega \cos(\omega t) p_z
\nonumber
\\
\label{eq:h3prime}
&+& A m \alpha \omega \cos(\omega t) \sigma_y. 
\eean
This result is useful: even in the range of
strong spin-orbit interaction $\tilde \alpha \sim \hbar \omega_0$
and strong driving $\tilde{E}_\text{ac} \sim \hbar \omega_0$, 
the qubit Hamiltonian obtained by projecting 
$H'''(t)$ to the lowest-energy two-dimensional
subspace provides an accurate description of the dynamics. 
Using the projector $Q = \ket{0 \uparrow} \bra{0 \uparrow}
+ \ket{0\downarrow} \bra{0 \downarrow}$, 
the qubit Hamiltonian reads
\bean
H'''_q(t) &=& Q H'''(t)  Q \nonumber
\\
&=& \label{eq:hq3prime}
- \frac{1}{2} \hbar \omega_\text{L} \sigma_z,
+ A m \alpha \omega \cos (\omega t) \sigma_y.
\eean
where the first, static term is the contribution from the
inhomogeneous Zeeman term $H'_\text{Z}(z)$ in 
Eq.~\eqref{eq:h3prime}, $\omega_\mathrm{L}$ is given by Eq.~\eqref{eq:hqubit},
and the second, driving term is the contribution from the
fourth term in Eq.~\eqref{eq:h3prime}.
Note that the arrows in the above definition of $Q$ do not
correspond to the physical spin degree of freedom, since
we have performed spin-dependent unitary transformations
along the way. 
Note also that the dynamics in the qubit subspace is coupled
to the higher-lying energy levels 
via $H'_\text{Z}(z)$ and the $p$-linear term in Eq.~\eqref{eq:h3prime},
but the corresponding coupling matrix elements are weak
compared to the level spacing $\hbar \omega_0$ 
as long as $\tilde{\alpha}, \tilde{E}_\text{ac} \lesssim \hbar \omega_0$
and the Zeeman splitting is small compared to the
orbital level spacing, $\omega_\text{L} \ll \omega_0$.

In conclusion, with our result
Eq.~\eqref{eq:hq3prime}, we have established that 
even in the nonperturbative regime
of strong spin-orbit interaction and strong driving, 
the qubit  can be described in an appropriately
chosen reference frame as a simple harmonically driven two-level
system.
Furthermore, a comparison of 
Eq.~\eqref{eq:hqubit} and \eqref{eq:hq3prime} reveals that
the analytical results for the qubit Larmor frequency 
is the same for the strong-driving case 
($\tilde{E}_\text{ac} \sim \hbar \omega_0$)
as for the weak-driving case
($\tilde{E}_\text{ac} \ll \hbar \omega_0$).
Even more, 
the Rabi frequency upon resonant
drive 
is also described by the same formula in the two cases: 
this is seen if the prefactor $A m \alpha \omega$ in 
Eq.~\eqref{eq:hq3prime} is evaluated 
using the resonance condition $\omega = \omega_\text{L}$
and expressed via the parameters $\tilde{E}_\text{ac}$, $\tilde{\alpha}$, 
and the resulting driving Hamiltonian is compared to
Eq.~\eqref{eq:hqubit}. 
Therefore, our analysis in this section have shown 
that the results obtained for the
weakly driven case can simply be extrapolated to the
strong-driving regime.

\section{Fast spin flip due to strong half-cycle excitation}
\label{sec:results}

\begin{figure}
 \centering
  \includegraphics[width=\columnwidth]{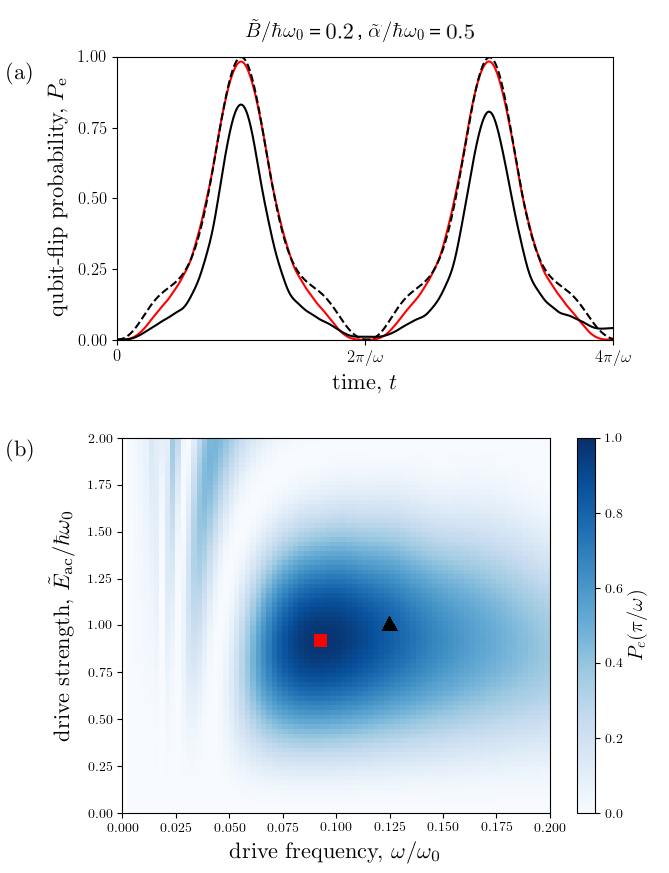}
  \caption{
  \label{fig:qubitflip}
  \textbf{
  Qubit-flip probability of a spin-orbit qubit
  due to a half-cycle electric excitation.}
  (a) Black solid line shows the
  numerical result for qubit-flip probability $P_\mathrm{e}(t)$, for
  drive frequency $\omega \approx 0.121 \omega_0$
  and
  $\tilde{E}_\mathrm{ac}=\hbar \omega_0$
  corresponding to the black triangle in (b). 
  For these parameters, the analytical result (dashed black line) 
  predicts a perfect qubit flip at $t=\pi/\omega$. 
  Red line: qubit-flip probability for parameter values
  $\omega \approx 0.092 \omega_0$ and
  $\tilde{E}_\mathrm{ac}= 0.916 \hbar \omega_0$,
  corresponding to the red square in (b). 
  (b) Numerical results for the qubit-flip probability after
  a half cycle, $P_\mathrm{e}(\pi/\omega)$,
  as a function of $\omega$ and $\tilde{E}_\mathrm{ac}$.
  Red square marks the maximum of the
  qubit-flip probability. 
  }
\end{figure}

In this section, we solve the time-dependent 
Schr\"odinger equation to 
investigate the accuracy of the analytical results
derived in the preceding section.

First, we outline the procedure to obtain numerical results
for the dynamics. 
The first step
is to expand the Hamiltonian 
\eqref{eq:h3prime} in the first $n_\text{max}$ 
harmonic oscillator eigenstates.
Due to the spin degree of freedom, 
the size of this truncated Hamiltonian matrix is 
$2n_\text{max} \times 2n_\text{max}$.
Second, the initial state in this time-dependent frame
is $\ket{\mathrm{g}'''}$,
which the state obtained via transforming 
the lab-frame ground state $\ket{\mathrm{g}}$ of the 
spin-orbit qubit with the three transformations
discussed above: 
$\ket{\mathrm{g}'''} =
S(t=0) W(t=0) U \ket{\mathrm{g}}$.
Note that both $S(t=0)$ and $W(t=0)$ are identity operators
(actually, they are identity operators for any
$t = N \pi/\omega$), 
which implies $\ket{\mathrm{g}'''} = \ket{\mathrm{g}'}$, hence we obtain
the initial state by diagonalizing the static part of $H'(t)$.
Third, we numerically solve the time-dependent Schr\"odinger 
equation governed by $H'''(t)$ to obtain the time evolution
$\ket{\psi'''(t)}$ of the initial state.
Fourth, we transform the solution back to the frame
of $H'$
via $\ket{\psi'(t)} = W^\dag(t) S^\dag(t) \ket{\psi'''(t)}$.
Fifth, we evaluate the qubit-flip probability via
$P_\mathrm{e}(t) = |\braket{\mathrm{e}'|\psi'(t)}|^2$.

Solving the time-dependent Schr\"odinger equation in 
the co-moving frame is computationally more efficient
than doing that in the lab frame. 
For the co-moving frame, as we increase the number $n_\text{max}$ 
of harmonic oscillator basis states, the qubit-flip
probability $P_\mathrm{e}(t)$ converges rapidly. 
For the parameter set discussed below, 
the choice 
$n_\text{max}=4$ already provides satisfactory convergence. 
To have the same accuracy from a lab-frame simulation, 
we need a significantly larger basis $n_\text{max} = 10$. 

In addition to the numerical approach, we can also describe
the driven qubit dynamics using an analytical approximation.
This is based on the effective qubit Hamiltonian $H'''_q(t)$ in 
Eq.~\eqref{eq:hq3prime}. 
We take the initial state as the ground state of the static
part of $H'''_q(t)$, assume that the dynamics described by
the wave function
$\ket{\psi'''(t)}$ is 
restricted to this two-dimensional qubit subspace, and 
within this subspace it follows the simple Rabi dynamics
according to the rotating-wave approximation. 
Again, the qubit-flip probability is obtained
via
$P_\mathrm{e}(t) = |\braket{\mathrm{e'}|W^\dag(t) S^\dag(t) \psi'''(t)|}|^2$,
where $\ket{\mathrm{e'}}$ is obtained perturbatively, as described 
in Section \ref{sec:perturbative}.

Here we will discuss certain features of the driven
qubit time evolution in 
the case of strong driving $\tilde{E}_\text{ac} \sim
\hbar \omega_0$,
when the 
amplitude $A = \frac{\sqrt{2} \tilde{E}_\text{ac}}{\hbar \omega_0} L $ 
of the real-space oscillations of the 
electronic wave function is comparable to the
width $L$ of the wave function. 
Note that a priori we do not expect that the numerical 
solution will show simple
Rabi-oscillation qubit dynamics under this strong-driving condition. 
In fact, it is known that even in simple magnetic resonance
with a spin-1/2, an increased driving strength leads to 
unconventional dynamics\cite{BlochSiegert,Shirley,romhanyi,Fuchs, PhysRevA.95.062321}.

Hence the first question we address is whether a high-quality
qubit-flip is achievable with a strong half-cycle excitation pulse?
Consider the example when the magnetic field
and the spin-orbit strength are set as
$\tilde{B} = 0.2 \hbar \omega_0$ 
and
$\tilde{\alpha} = 0.5 \hbar \omega_0$. 
For this setting, the  results
\eqref{eq:larmor},
\eqref{eq:tau}, and \eqref{eq:naiveteac} suggest
that the fastest qubit-flip is achievable by a strong half-cycle 
pulse ($N=1$) with resonant drive frequency 
$\omega = \omega_L \approx 0.121 \omega_0$ 
and drive strength
$\tilde{E}_\text{ac} = \hbar \omega_0$.
The
analytical (numerical) time evolution of the 
qubit-flip probability $P_\mathrm{e}(t)$ is shown for
this parameter set in 
Fig.\ref{fig:qubitflip}a as a dashed (solid) black line.
On the one hand, the comparison of the two results
reveals that the actual dynamics (solid) deviates
from the analytical approximation (dashed). 
On the other hand, we also see that a fairly 
high qubit-flip probability $P_\text{e} \approx0.8$
is achieved at $t=\pi/\omega$.

The latter observation suggests that a high-fidelity
qubit flip may be achieved with a strong half-cycle
pulse by slightly adjusting the drive parameters.
To explore this possibility, we extend the numerical 
simulations of the qubit dynamics to 
a range of values of drive frequency $\omega$
and drive strength $\tilde{E}_\text{ac}$. 
We plot the qubit-flip probability $P_\mathrm{e}(\pi/\omega)$
at the end of a half-cycle excitation  
as a function of these parameters
in Fig.~\ref{fig:qubitflip}b; there
the black triangle corresponds to the black solid curve 
of Fig.~\ref{fig:qubitflip}a discussed above. 
In Fig.~\ref{fig:qubitflip}b, the highest qubit-flip probability
is found in the point marked with the red square,
where $P_\mathrm{e}(\pi/\omega) \approx 0.98$. 
The corresponding complete time evolution $P_\text{e}(t)$ 
is shown as the red curve in Fig.~\ref{fig:qubitflip}a.
These results demonstrate that even though the 
qubit-flip probability $P_\text{e}(t)$ 
is more complex than a simple Rabi oscillation
\eqref{eq:rabiosc} in case 
of a strong half-cycle drive, a high-fidelity qubit flip can still be
reached by fine tuning the parameters of the drive. 

\begin{figure}
\includegraphics[width=\columnwidth]{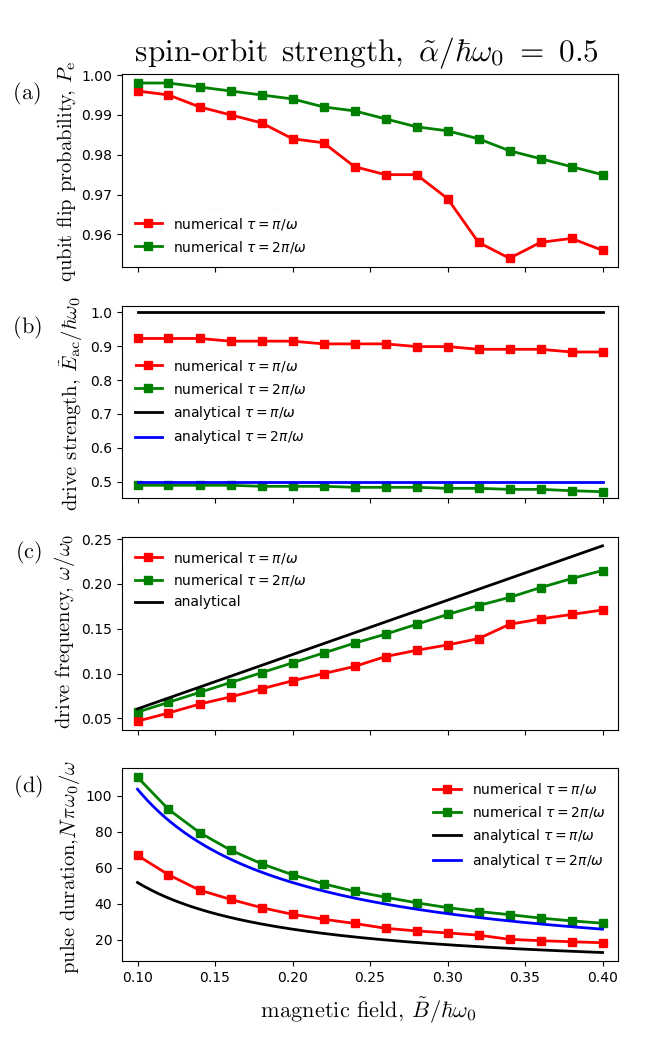}
\caption{
\textbf{
Magnetic field dependence of qubit-flip 
pulse parameters.}
(a) Numerically computed 
maximal qubit-flip probability after a half-cycle (red)
or single-cycle (green) drive pulse. 
(b,c,d) The drive strength $\tilde{E}_\mathrm{ac}$, 
drive frequency $\omega$,
and pulse duration $\tau$, corresponding to the
maxima of $P_\mathrm{e}$ shown in (a).
Solid lines show the analytical results.
\label{fig:max}}
\end{figure}

Recall that the considerations in Sec.~\ref{sec:estimates},
especially Eq.~\eqref{eq:tau}, suggest that 
the qubit flip can be made faster by increasing the 
Larmor frequency via increasing the magnetic field. 
To test this expectation, we extend the numerical investigation 
to a range of magnetic field values, and report the 
results in Fig.~\ref{fig:max}.
For each value of the magnetic field, 
we compute the qubit-flip probability 
$P_\text{e}(\pi/\omega;\omega,\tilde{E}_\text{ac})$
upon a half-cycle pulse, and maximize
$P_\text{e}(\pi/\omega)$ 
with respect to $\omega$ and $\tilde{E}_\text{ac}$.
This maximum point for a certain magnetic-field value
was shown as the red square in Fig.~\ref{fig:qubitflip}b.
Red squares in 
Fig.~\ref{fig:max}a show the maximal qubit-flip probability 
found by this procedure as function
of magnetic field, and in Figs.~\ref{fig:max}b and c,
the drive strength $\tilde{E}_\text{ac}$ and drive frequency
$\omega$ corresponding to this maximum.
In Fig.~\ref{fig:max}b and c the black solid line shows
the analytical result Eq.~\eqref{eq:naiveteac} and \eqref{eq:larmor}
respectively.
In Fig.~\ref{fig:max}d,  red squares and black line 
show the pulse duration $\tau = \pi / \omega$, derived directly from 
the data in Fig.~\ref{fig:max}c.

The main conclusions drawn from Fig.~\ref{fig:max}
are as follows. 
(i) Fig.~\ref{fig:max}d confirms the expectation that 
a larger magnetic field allows for a faster qubit flip. 
(ii) Even though faster qubit flips can be achieved 
by increasing the magnetic field, the numerical 
data (red squares) in Fig.~\ref{fig:max}b
confirms that 
the drive strength does not have to be increased to achieve that. 
(iii) The qubit flips are not perfect: 
Fig.~\ref{fig:max}a shows that the larger the magnetic field, 
the smaller the maximal qubit-flip probability.
This is important if this qubit-flip process is considered
in the context of quantum information processing: 
the result suggests that by increasing the magnetic field, 
a faster, but less precise quantum-logical operation is realized.
(iv) The trends and orders of magnitude 
predicted by the estimates in 
Sec.~\ref{sec:estimates} (black solid lines in Fig.~\ref{fig:max}) 
are confirmed by the numerical data (red squares in Fig.~\ref{fig:max}),
although there are quantitative deviations of a few tens of percents.

Experimental constraints might inhibit the application of 
a half-cycle pulse: it might be challenging to produce
such a strong electric field, or if realized, the strong field could
ionize the quantum dot. 
This motivates to consider the case when a drive pulse with the
duration of a few half-cycles is applied. 
For example, Eq.~\eqref{eq:naiveteac} suggests that
the drive strength required for 
a single-cycle pulse ($N=2$, $\tau = 2 \pi/\omega$) is half of 
the drive strength of the half-cycle pulse; 
this is shown as the blue line in Fig.~\ref{fig:max}b. 
The corresponding numerical result (green squares)
confirms this. 
Note that Fig.~\ref{fig:max}a shows that the precision 
of the qubit-flip improves by using a single-cycle pulse
instead of a half-cycle one.
This is expected, as the qubit dynamics is expected to approach
the simple Rabi dynamics as the drive strength is reduced. 
Finally, since the half-cycle and single-cycle pulses are both
approximately resonant with the qubit Larmor frequency, 
the latter one is half as fast as the former one, 
which is also seen in Fig.~\ref{fig:max}d. 

Our results discussed in this section are all phrased in terms
of dimensionless parameters; here we 
make connection to realistic parameter values. 
Again we take the example of an InAs quantum dot, with
parameters specified in Sec.~\ref{sec:estimates}.
Then the dimensionless parameters $\tilde{B}/\hbar \omega_0=0.2$,
$\tilde{E}_\mathrm{ac}/\hbar \omega_0=0.916$, $\omega/\omega_0=0.092$
corresponding to the red square in Fig.~\ref{fig:qubitflip}
are translated to magnetic field $B=0.35$ T, drive strength 
$E_\mathrm{ac}=22$ kV/m, (or oscillation amplitude $A=75$ nm) 
and drive frequency $\omega/2\pi=22$ GHz.

\section{Discussion and conclusions}
\label{sec:discussion}

\emph{Effect of the spin-orbit strength on the
qubit-flip time.}
In Sec.~\ref{sec:results}, we considered a fixed spin-orbit strength
$\tilde{\alpha}$, and concluded that the qubit-flip time
can be shortened by increasing the magnetic field.
However, the spin-orbit strength is also tunable
experimentally \cite{PhysRevB.79.121311, PhysRevB.91.201413,doi:10.1021/nl301325h, scherubl-electrical_tuning}.
Therefore, 
a complementary question concerns the case of a fixed magnetic
field: how does the qubit-flip time depend on the strength
of the spin-orbit interaction?
In the absence of any practical constraints, a simple and slightly 
counterintuitive answer can be based on 
the naive estimates in Sec.~\ref{sec:estimates}:
the qubit-flip time can be reduced by reducing the spin orbit strength.
This is a consequence of the fact that in our model, the qubit
Larmor frequency increases when the spin-orbit strength
is reduced (see Eq.~\eqref{eq:larmor}), and therefore the 
duration of a resonant half-cycle pulse gets shorter. 
Our numerical simulation results (not shown) do 
confirm this expectation. 
A practical constraint is the following. 
By reducing the spin-orbit strength, the drive strength 
$\tilde{E}_\text{ac}$ required for the qubit flip 
grows, according to Eq.~\eqref{eq:naiveteac}.
In an actual device, the drive strength corresponding to 
the ionization of the quantum dot cannot be exceeded, which
implies a minimal value of the spin-orbit strength.
That minimal spin-orbit strength provides a lower bound 
on the qubit flip time, and there is no practical advantage
of going below that minimal spin-orbit strength.

%

\emph{Further directions.}
We have been focusing on the gross
features of the dynamics of spin-orbit qubits 
induced by strong electric fields. 
A further, more refined analysis is motivated by the potential
application of this setup in quantum information processing; 
that should incorporate a more generic
description of the spin-orbit interaction,
the investigation of 
quantum-gate fidelities, realistic and optimized
electric pulse shapes beyond the harmonic-driving paradigm,
as well as decoherence processes.
We think that the numerical and analytical methods we
developed here, based on the co-moving frame transformation,
will serve as useful tools for addressing such extensions. 

\emph{Conclusions.}
In this work, we have described the electrically driven 
dynamics of a spin-orbit qubit, in the regime of strong
driving, when the ampiltude of spatial oscillations of the
electron are comparable to the width of its wave function. 
Using the model of a one-dimensional quantum dot
subject to Rashba spin-orbit interaction, 
we have demonstrated that strong electric pulses with
short duration allow for fast, high-precision qubit control,
potentially on the sub-nanosecond time scale. 
Our analysis of the dependence of the shortest achievable
qubit-flip time scale on experimentally tunable 
parameters (magnetic field, spin-orbit strength)
offers practical guidelines to optimize electrically 
induced single-qubit operations on spin-orbit qubits.

\acknowledgments
We thank P. Boross and G. Sz\'echenyi for useful feedback on 
the manuscript. 
This	research	was	supported	by	the	
National	Research	Development	and	
Innovation	Office	of	Hungary	within	the	Quantum	
Technology	National	Excellence	
Program		(Project	No.	2017-1.2.1-NKP-2017-00001).
AP acknowledges funding from the NKFIH Grants
105149 and 
124723, and the New National Excellence Program of 
the 
Ministry of Human Capacities of Hungary. 
 
\appendix

%

%

\bibliography{forras}
\bibliographystyle{ieeetr}
\end{document}